\documentclass[12pt, a4paper]{memoir} 
\usepackage[french,english]{babel}

\usepackage [vscale=0.76,includehead]{geometry}                

\usepackage{lipsum}
\usepackage{graphicx}
\usepackage{amsmath}
\usepackage{fullpage}
\usepackage{mathptmx} 
\usepackage{helvet} 
\usepackage{relsize}
\usepackage[T1]{fontenc}
\usepackage{tikz}
\usepackage{booktabs}
\usepackage{textcomp}
\usepackage{multirow}
\usepackage{pgfplots}
\usepackage{url}
\usepackage{footnote}
\usepackage{csquotes}
\usepackage{xcolor}
\usepackage{amsmath}
\usepackage{algorithm}
\usepackage{algpseudocode}
\usepackage{amssymb}
\usepackage{paralist}
\usepackage{bbm}
\usepackage{stmaryrd}
\usepackage{diagbox}
\usepackage{multirow}
\usepackage{graphicx}
\usepackage{amsfonts}
\usepackage{mathtools}
\usepackage{xspace}
\usepackage{amsthm}
\usepackage{colortbl}
\usepackage{verbatim}
\usepackage{hyperref}
\usepackage{amssymb}
\usepackage{tikz}
\usepackage{scalerel}
\usepackage{pict2e}
\usepackage{pgfplots}
\usepackage{enumitem}
\usepackage[usestackEOL]{stackengine}
\usetikzlibrary{arrows,calc}
\usepackage{chngcntr}

\usetikzlibrary{arrows,shapes,positioning,shadows,trees}
\makesavenoteenv{tabular}
\makesavenoteenv{table}

\headstyles{komalike}
\nouppercaseheads
\chapterstyle{dash}
\makeevenhead{headings}{\sffamily\thepage}{}{\sffamily\leftmark} 
\makeoddhead{headings}{\sffamily\rightmark}{}{\sffamily\thepage}
\makeoddfoot{plain}{}{}{} 
\makeheadrule{headings}{\textwidth}{\normalrulethickness}

\setsecnumdepth{subsection}
\newcommand \somme[3] {\sum\limits_{\substack{#1}} ^{#2} #3}
\newcommand{\DO}{\mathsf{DO}}
\newcommand{\Items}{\Psi}
\newcommand{\Users}{\Upsilon}
\newcommand{\itemIndex}{\mathsf{i}}
\newcommand{\userIndex}{\mathsf{u}}
\newcommand{\Item}{\psi}
\newcommand{\User}{\upsilon}
\newcommand{\ItemRepr}[2]{\Item ^{#2}_{#1}}
\newcommand{\UserRepr}[2]{\User ^ {#2}_{#1}}

\newcommand{\COS}{\mathsf{COS}}

\newcommand{\loss}{l}
\newcommand{\gradientLoss}{\partial \loss}
\newcommand{\ELoss}{\mathcal{L}_{B_u}}
\newcommand{\userInteract}{C}
\newcommand{\Block}[3]{B^{#1, #2}_#3}
\newcommand{\Positive}[2]{\Pi_#1^{#2}}
\newcommand{\Negative}[2]{\eta_#1^{#2}}
\newcommand\proto{\textsc{DRIFT}}

\pretitle{\HUGE\sffamily \bfseries\begin{center}} 
\posttitle{\end{center}}
\preauthor{\LARGE  \sffamily \bfseries\begin{center}}
\postauthor{\par\end{center}}
\newcommand{\jury}[1]{%
\gdef\juryB{#1}} 
\newcommand{\juryB}{} 
\newcommand{\session}[1]{%
\gdef\sessionB{#1}} 
\newcommand{\sessionB}{} 
\newcommand{\option}[1]{%
\gdef\optionB{#1}} 
\newcommand{\optionB} {}

\option{Type your Option Here} 
\title{DRIFT: A Federated Recommender System with Implicit Feedback on the Items}
\author{Theo Nommay}
\date{} 
\jury{
Research project performed at YOUR LAB \\\medskip
Under the supervision of:\\
Massih-Reza Amini\\Radu Ciucanu\\Marta Soare\medskip
Defended before a jury composed of:\\
Massih-Reza Amini\\
Philippe Mulhem\\
Georges Qu\'enot\\
Marta Soare\\
Danielle Ziebelin
}
\session{September \hfill 2022}

\begin{document}
\selectlanguage{English} 
\frontmatter
\begin{titlingpage}
\maketitle
\end{titlingpage}

\setlength{\parskip}{-1pt plus 1pt}

\renewcommand{\abstracttextfont}{\normalfont}
\abstractintoc
\begin{abstract}

Nowadays there are more and more items available online, this makes it hard
for users to find items that they like. Recommender systems aim to find the item
who best suits the user, using his historical interactions. Depending on the context,
these interactions may be more or less sensitive and collecting them brings an
important problem concerning the users’ privacy. Federated systems have shown
that it is possible to make accurate and efficient recommendations without storing
users’ personal information. However, these systems use instantaneous feedback
from the user. In this report, we propose DRIFT, a federated architecture for recommender
systems, using implicit feedback. Our learning model is based on a
recent algorithm for recommendation with implicit feedbacks SAROS \cite{ABCILM21}. We aim
to make recommendations as precise as SAROS, without compromising the users’
privacy. In this report we show that thanks to our experiments, but also thanks to a
theoretical analysis on the convergence. We have shown also that the computation
time has a linear complexity with respect to the number of interactions made. Finally,
we have shown that our algorithm is secure, and participants in our federated
system cannot guess the interactions made by the user, except DOs that have the
item involved in the interaction.

\end{abstract} 
\abstractintoc

\renewcommand\abstractname{Acknowledgement}
\begin{abstract}
I would like to express my sincere gratitude to Marta Soare, Radu Ciucanu,
and Massih-Reza Amini for their invaluable assistance throughout the internship.
I would also like to express my sincere thanks to Aleksandra Burashnikova, for all
the help she gave me in understanding her algorithm.
\end{abstract}

\cleardoublepage

\tableofcontents* 
\normalsize

\mainmatter
\SingleSpace

\counterwithout{figure}{chapter}
\counterwithout{table}{chapter}

\begin{table}[t]

\begin{tabular}{|c|c|}
\hline
I & The total number of items
\\ \hline
U & The total number of users
\\ \hline
$\itemIndex$ & The index of the item i
\\ \hline
$\userIndex$ & The index of the user u
\\ \hline
$\Items$ & The set of item representation
\\ \hline
$\Users$ & The set of user representation
\\ \hline
$\ItemRepr{i}{t}$ & The d-dimension vector who represent the item i at timestamp t
\\ \hline
$\UserRepr{u}{t}$ & The d-dimension vector who represent the user u at timestamp t
\\ \hline
$S_u$ & The scores of items for user u
\\ \hline
\hline
$\DO_i$ & The Data Owner i 
\\ \hline
$\COS$ & The Central Orchestration Server 
\\ \hline
\hline
$\userInteract = (u, i, is\_positive) $ & The triplet representing the interaction of the user u with the item i. 
\\ \hline
$\Block{k}{t}{u}$ & The block for the user u at timestamp t created by $\DO_k$ 
\\ \hline
$\Positive{u}{t}$ & The list of positive interaction in the block for the user u at timestamp t
\\ \hline
$\Negative{u}{t}$ & The list of negative interaction in the block for the user u at timestamp t
\\ \hline
\hline
$ \loss_{\userIndex,\itemIndex,\itemIndex'}$ & The loss for user u and two items i and i'
\\ \hline
$\ELoss $ & Empirical ranking loss with respect to a block of items
\\ \hline
$\sigma $ & The sigmoid function
\\ \hline
\end{tabular}
\caption{All the notation used}\label{Theorical:notation}
\end{table}

\chapter{Introduction}
\section{Problem statement}
With the increasing number of available items, recommender systems have more and more
interactions to manage. These interactions are usually treated with a collaborative filtering
algorithm, who needs to collect and store all the data in one central server. However, it is
not always possible to collect user data because of potentially sensitive data, for example in
the medical or financial context. Some organizations may profit from having access to these
data, by selling them or merging them with other organizations, which may jeopardize users’
privacy. To solve this problem, recent works use federated learning for their recommender
system. A federated recommender system aims to update the learning model without sending
the user interaction. To do that, it computes the values needed to update separately from the
central server, to hide potentially sensitive information. These systems mostly take only into
account a positive interaction of the user, with matrix factorisation or reinforcement learning
for example. This is problematic because it does not take into account when the user has no
interaction with the proposed item. Moreover, it does not keep the context of the interaction to
understand the choice of the user. For example a user may choose an article among those that
are proposed, but other items that are not proposed may please him even more.
\section{Scientific Approach and Investigative Method and Results}
Using implicit feedback, our objective is twofold
\begin{itemize}
\item We aim to build an efficient federated architecture which is able to recommend items
that best suits the user. A federated architecture is based on multiple Data Owners (DOs)
that are orchestrated by a Central Orchestration Server (COS). Updating a model in a
federated architecture consist to hide the user’s information from the central server. To
do that, all the computations needed with these informations are computed locally on
the users devices. The organizations containing the items that users interact with, are
represented as different DOs. We summarize this principles in the Figure 2.2.
\item We also decide to guarantee data security by avoiding any leak in the architecture, using
cryptographic tools.
\end{itemize}

Our federated architecture is orchestrated by the COS. Each item is stocked into multiples
DOs, who can communicate with both users and COS. Our learning algorithm is built following
a recent recommender system with implicit feedback, SAROS \cite{ABCILM21}. Each DO will store the
user’s negative and positive interaction, as a block of interaction. When a block is complete the
DO will communicate with the COS to update the learning model.
The objective of our algorithm is threefold, we must ensure relevant recommendations, we
must ensure a low computation time, and we must ensure the security of the users. In this
report :

\begin{itemize}
\item We analyze the precision of our recommendation, using the MAP and NDCG measure.
Moreover, we provide a proof of convergence, which shows that the loss of our algorithm
will converge to the same loss as a non secured version, in other words, that the relevance
of the recommendation will be similar to a non secured algorithm.
\item We analyze the complexity of our algorithm thanks to a theoretical analysis, where we
have shown that our algorithm has a linear complexity in the number of interactions.
Moreover, our experiments have shown that more than 90\% of the time is dedicated to
the update of the model.
\item We provide theoretical analysis where we have shown that all the participants of our federated
system are unable to guess the users’ interactions, even if they intercept messages
in the system.
\end{itemize}

\section{Contents of this report}
The rest of this report is organized as follows.
Chapter 2 presents related works, we start by presenting the context in Section \ref{sec:21}. Our context
and motivation comes from the last recsys’ 21 conference, where the privacy of the user
was a crucial point, and where a lot of secure recommender systems were based on federated
learning. For this reason we will next focus on Federated Learning in Section \ref{sec:22}, where we are
positioning our algorithm, with respect to related works. Then we will set the environment in
Section \ref{sec:23}, we have chosen to set the participants honest-but-curious. After the architecture is
decided, we focus on recommender systems on Section \ref{sec:25} where we will see in detail different
recent federated and non federated recommender systems with implicit and explicit feedback.
As we said in Section 1.2, we set our learning algorithm following the SAROS architecture.
Since we have set our participants honest but curious, we need to secure the communication
between the participants, to do that we have to relate to encryption tools that we present in
Section \ref{sec:24}.

Chapter 3 presents our algorithm, we start by presenting what happens at the initialization
of the algorithm in Section \ref{sec:31}. We will then focus on the recommendation to the user in Section
\ref{sec:32}. Then we present the core of our algorithm in Section \ref{sec:33}, where we present how DOs
build the local parameters thanks to user interactions. We finish with Section \ref{sec:34}, where we
present how the COS updates the global parameters thanks to the local parameters built in the
DOs.
Chapter 4 presents the technical choices that we made in our algorithm. We start by presenting
the repartition of the task in Section \ref{sec:41}, where we explain different possibilities with
their advantages and their drawbacks. We finally chose to compute gradients directly in the DO
and to share them with the COS. We will then focus on the computations of these gradients in
Section \ref{sec:42}. We finish with the encryption of the sensitive information.
Chapter 5 presents the technical analysis that we provide. We start by showing that our
algorithm can converge to the same minimizer as a non secure algorithm in Section \ref{sec:51}. Then
we analyze the security of our framework in Section \ref{sec:53}. Our objective here is to show that
all participants cannot guess the users’ interaction with a probability better than random, even
if they intercept messages in the network. We finish with Section 5.3 where we compute the
complexity of DRIFT. We have shown that the complexity of our secure algorithm is linear in
terms of the number of interactions.
Chapter 6 presents the experiments that we made. We aim to have recommendations as
relevant to SAROS. In Section \ref{sec:61} we present the metrics that we used to compare these two
algorithms, and all the settings of our experiments In Section \ref{sec:62} we analyze the results of our
recommendation, we do it in two parts. We first analyze the evolution of the loss, where we
saw that DRIFT converges earlier than SAROS, then we analyze the evolution of the metrics s
after DRIFT converges and after SAROS converges. This part confirm the assumption of
convergence made in Chapter \ref{chap:5}.

\chapter{State-of-the-art}

In this Chapter we will see the current scientific state of the art about security in recommender
systems. To do that we will first focus on the motivation and context of the work in Section .
After that, we will see in detail what has been done in federated learning, in Section , and how
we can encrypt the messages in Section . After that, we will focus on recents Recommender
Systems in Section .

\section{Context}
\label{sec:21}
Federated Learning is a machine learning technique that trains a model across multiple service
providers while keeping the training data locally. This technique is used to avoid a centralized
system, due to potential sensitive data. Moreover, as said in Introduction, there is an urge to
focus on security in recommender systems. This motivates us to focus on security in recommender
systems, using the federated learning paradigm. Moreover, recent works as \cite{Ciucanu} shows
that it is possible to have a relevant algorithm with a secure framework, thanks to federated
learning.
In addition we saw in the last Recsys21 conference\footnote{\url{https://recsys.acm.org/recsys21/}}, where the privacy of users in recommender
systems was a crucial point. A lot of these works show that the collecting and handling
potentially sensitive data raises serious privacy problems, as \cite{TB2021} which show that if there is
a collaboration between different vendors to increase the size of their dataset, it poses an important
privacy problem for vendors and users. To avoid a central server, recents works use a
federated system, as \cite{LMBH2021} who show that federated method is better suited to protect privacy.
Moreover some other recent works transform the most used algorithm into a federated one to
avoid a central server. For example \cite{WangHPW20} who create a federated system for recommendation
collaborative filtering, or \cite{ChaiWCY21} who create a federated system for recommendation with matrix
factorization. This motivates us to build a Federated architecture to solve our problem.

\section{Federated learning}
\label{sec:22}
We aim to use this technique to build a secure learning system, this is why we built DRIFT
upon the typical federated learning characteristics. First, when a DO manages interactions,
local parameters are created. The DO keeps them locally, and each one has only access to his own data. Since the COS orchestrates the entire system, and can communicate with all DOs, it
must manage the global parameters. The data of an organization will be stored into a dedicated
DO. 
\cite{ZWZ2021} notice that the informations of the score given by a user shall also be well protected, and
that the server should not know which user interacted with which item. So that COS does not
know which user is updating its values, we will store the user interactions into different DOs,
who can communicate with the COS. Recent works in Federated Recommender Systems, store
the information of the user directly on their devices, as \cite{DBLP:conf/emnlp/QiWWH020} which save the historical clic news
in it. In our case, since we will need different DO to save the items, we will prefer a crosssilo
architecture, where different DO will store the user information. We are positioning our
architecture following a recent cross-silo algorithm, SAMBA \cite{Ciucanu}, where we keep some typical
characteristics:
\begin{itemize}
\item \textbf{Distribution scale.} We have few DO, generally less than 100. They are created for each
organization in the system, thanks to clustering.
\item \textbf{Addressability.} The COS can access a specific DO thanks to an id.
\item \textbf{Statefulness.} Each DO maintains local variables throughout the execution of the entire
algorithm.
\item \textbf{Data partition.} The partition should be fixed. In our case, we assume that the number
of organizations in our system is fixed, which implies that the data in each DO is fixed.
\item \textbf{Incentive mechanisms.} To ensure the honest participation of each DO we need incentive
mechanisms. In our case, since each DO is related to a unique organization, we assume
that they are business competitors, with a monetary gain according to the number of items
chosen by the user.
\end{itemize}
\section{Environment}
\label{sec:23}
Each participant in our environment can be a user, a COS or a DO. They can be
\begin{itemize}
\item Untrusted, all the participants will always send a result that suits them, even if it is a bad
one, and will try to analyze our data.
\item Honest but curious, all the participants will send good results but will try to analyze our
data.
\item Honest, all the participants will send good results and will never try to analyze our data.
\end{itemize}
In our case, the knowledge of our participants is limited and totally independent, so it is not
a problem if they are trying to analyze the data. But we need to be sure that the results that they
send are correct, so we will suppose that they are honest but curious.

\section{Encryption}
\label{sec:24}
Since we supposed that our participants are honest but curious, they can sniff the network and
intercept messages sent in our algorithm. This must not compromise the privacy of the user.
To avoid that, we need to encrypt the messages, thanks to cryptographic tools. As we need
to have an efficient algorithm, DRIFT relies on a NIST standard for crypto-system, AESGCM
\cite{AES-FIPS}. The AES-GCM crypto-system is defined by a triplet of polynomial-time algorithms
(Gen,Enc,Dec) and a security parameter $\lambda$ such that Gen($1\lambda$) generates a uniformly random
symmetric key, according to $\lambda$. Let $c_1 = Enc(m_1)$ be the encryption of a message $m_1$ and let
a message $m_2 = Dec(c_1)$ be the decryption of $c_1$. If the symmetric key $K$ is the same for both
operations, then $m_1 = m_2$. In our case we need to hold this assumption, so we create keys at
the very beginning, and we share to each DO his private key, when they enter into the system.
Then all of these participants can read and write encrypted information. Moreover AES-GCM
is IND-CPA secure \cite{BellareDJR97}, so if an external attacker can read an encrypted message, he cannot
guess the decrypted one with a probability better than random.

\section{Recommender Systems}
\label{sec:25}
When a user interact with a recommender system, it creates an interaction. It can be
\begin{itemize}
\item \textbf{Explicit,} where the user gives a score to the item.
\item \textbf{Implicit,} where the feedback depends on if the user click \cite{Amini00}.
\end{itemize}
It is harder to manage implicits feedback, if a user does not click on an item, it does not mean
that the user is not interested. On the other hand, if a user clicks on the item, it does not mean
that the recommendation is pertinent, maybe she wants to click on it anyway. But since it is easier
to collect implicit feedback (generally in the form of clicks), we adapt DRIFT for implicit
rewards. There are multiple ways to update the model of a federated recommender system,
recents works are mainly based on Matrix factorization as \cite{DBLP:conf/sigir/LinRCRY0RC20} where the model is updated
thanks to a MLP on the rating of the user. A lot of works are also based on reinforcement learning,
as \cite{DBLP:conf/isit/LiSF20}, which use UCB in their federated recommender system. But all these approaches
mainly consider positive interactions, and the main hypothesis of \cite{ABCILM21} is that the user preference
is better represented when we consider a local sequence of positive and negative interactions on
the items. This is why we position our algorithm following the learning characteristics of their
algorithm SAROS. It consists of creating one block of items per user, when a positive interaction
follows a negative one, the block is complete and the model parameters are updated. We
aim to make recommendations as precise as SAROS, without compromising the users’ privacy.
We will consider that the COS stores the model parameters, and each DO will create the blocks
of interaction, and update these parameters.

\chapter{Algorithm DRIFT}
DRIFT has 4 different principal steps:
\begin{itemize}
\item The Initialization takes place before any user enter in the system, here we initialize DOs
and the COS global parameters. We present it in Section \ref{sec:31}.
\item The Recommendation takes place after a user enters in the system, here the COS, that
has global parameters, will return to the user the item that is supposed to please him the
most. We present it in Section \ref{sec:32}.
\item The Core of our Algorithm takes place while a user interact with items, here we DOs are
updating locals parameters. We present it in Section \ref{sec:33}.
\item The Update of Global Parameters takes place when a DO has complete blocks of interactions,
where it will share information with the COS. We present it in Section \ref{sec:34}.
\end{itemize}
\section{Initialization}
\label{sec:31}
We represent each item and user as a d-dimension vector, we denote $ \ItemRepr{\itemIndex}{t}$ (resp. $\UserRepr{\userIndex}{t}$) the value of the item (resp. user) at the index  $\itemIndex$ (resp. $\userIndex$), at time t. When the system is initialize, the $\COS$ will create vectors $\Users \in R ^ {d \times U} $ and $\Items \in R ^ {d \times I} $, who contains all the user and item representation. At this point, the $\COS$ has no information about the user and items, so it will initialize these vector at random. When an organization O enter in the system, it will give us all his items $\Items_\mathsf{O}$. To split these data into different $\DO$, we will use a non supervised algorithm to create clusters on this set of items. Each cluster of items $\Items_{\mathsf{O},k} \subseteq  \Items_\mathsf{O} $ will be assign to a $\DO$. If an item is present in multiple clusters, it will be assigned to multiple $\DO$.  

\section{Recommendation}
\label{sec:32}
The principal objective is to make recommendations, to do that we need to learn the user and the items, depending on his interactions.  The score of an item $\itemIndex$ for the user $\userIndex$ will be equal to the dot product between they two representation vector. To recommend item to a user, the $\COS$ will first create the score vector $S_\userIndex \in R ^ {|\Items|}  = \Items \times \Users \left[\userIndex \right] = \Items \times \UserRepr{\userIndex}{t} $, which contains the score of all the items for the user u. The $\COS$ will then return to the user, the item with the highest score.

\begin{figure}
\begin{center}
\includegraphics[width=.9\textwidth]{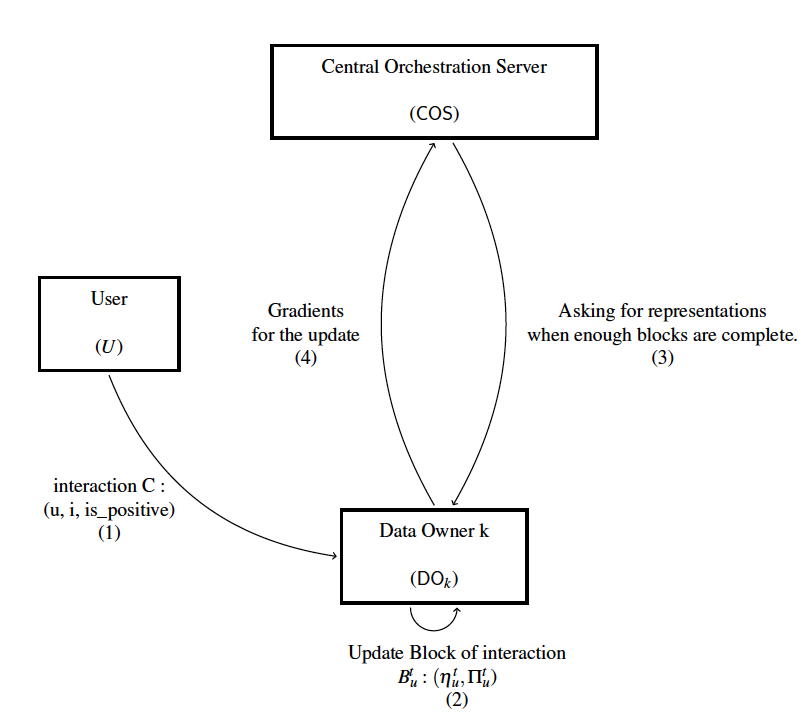}
\end{center}
\caption{The Workflow of The Algorithm}\label{annex:scheme}
\end{figure}

\section{Core of the Algorithm}
\label{sec:33}
The core of our algorithm is composed in 4 steps, 
\begin{itemize}
\item Manage the user interaction, 
\item Update the local parameters, 
\item Asking representations for computing the gradients;
\item Update the global parameters. 
\end{itemize}
We schematized the workflow of $\proto$ in Figure \ref{annex:scheme}. We assume that the $\COS$ and $\DO$ are well initialized, following the steps described in Section \ref{sec:31}. 

The step (1) begins after a user interact with an item. This interaction is represented as $\userInteract$, who contain the user id, the item id and the type of interaction (since we are in implicit context, it can be positive or not). Thanks to his database, the $\COS$ can find all the $\DO$ where the item is contained. 
These $\DO$ receive the $\userInteract$, and create the user block $ \Block{k}{t}{\userIndex}$. This block is composed of 2 vectors of items, the positive interactions $\Pi$ and negative interactions $\eta$. \\

The step (2) begins after a $\DO$ receives an interaction, and has to update his block. There are 3 state possible : 
\begin{enumerate}
\item When it receives a positive interaction, it closes the vector $\Negative{\userIndex}{t}$, and adds it to $\Positive{\userIndex}{t}$.
\item When it receives a negative interaction, it adds it to the vector $\Negative{\userIndex}{t}$.
\item When it receives a negative interaction while the vector $\Negative{\userIndex}{t}$ is close, it considers the block as complete, and starts filling a new block with this interaction.
\end{enumerate}
The principal advantage of this block is that it keeps the context of the interactions, by saving all negative interactions before the positive one.

The steps (3) and (4)  begin after a block is completed. Here we need to update the global parameters. To do that the $\DO$ asks the $\COS$ the representation of the user and the items in $ \Block{k}{t}{\userIndex}$. Thanks to these representations, the participants are able to update the global parameters. We give more details below.

\section{Update of Global Parameters}
\label{sec:34}
We aim to make recommendations by keeping the context of the interactions. If the user $\userIndex$ chooses the item  $\itemIndex$ over $\itemIndex '$ the score of  $\itemIndex$ should be higher than the score of  $\itemIndex '$ for this user. The goal is to avoid this miss ranking over all the items and the set of users, in other words we want to avoid that 
$S_\userIndex\left[\itemIndex\right] < S_\userIndex\left[\itemIndex '\right] 
\Leftrightarrow
\Items \left[\itemIndex \right] \times \Users \left[\userIndex \right] < \Items \left[\itemIndex '\right] \times \Users \left[ \userIndex \right] 
\Leftrightarrow
\Users \left[ \userIndex \right] \times ( \Items \left[\itemIndex \right] -  \Items \left[\itemIndex '\right]) < 0 
\Leftrightarrow
\UserRepr{u}{} \times ( \ItemRepr{i}{} -  \ItemRepr{i'}{} ) < 0$ . 
We denote $w = \UserRepr{u}{} \times ( \ItemRepr{i}{} -  \ItemRepr{i'}{} )$ and $\loss_{\userIndex,\itemIndex,\itemIndex ',}(w)$ the ranking loss for the items  $\itemIndex$,   $\itemIndex '$  and the user u. The result of the loss should be low when w is high, and high when w is low, so we aim to minimize this loss. To do that we will use the gradient descent principle. The idea is to update the global parameters with the opposite of the loss gradient, this will make the loss tend towards 0. It is possible if and only if the loss is differentiable (it is possible to derive it at any point of the domain). In our case the loss depends on 3 representations, the user, the item  $\itemIndex$ and the item  $\itemIndex '$. The gradient will be the vector

$$ 
\nabla \loss_{\userIndex,\itemIndex,\itemIndex'}(w) = \left[\begin{array}{c}
\dfrac{\gradientLoss_{\userIndex,\itemIndex,\itemIndex ',}}{\partial \UserRepr{u}{}}(w)\\ \\
\dfrac{\gradientLoss_{\userIndex,\itemIndex,\itemIndex ',}}{\partial \ItemRepr{i}{}}(w) \\ \\
\dfrac{\gradientLoss_{\userIndex,\itemIndex,\itemIndex 'm,}}{\partial \ItemRepr{i'}{}}(w)
\end{array}\right]
$$

As shown in Section \ref{sec:contrib:core}, the update of the model takes place after completing a local block on interaction. We need to loop over all negative items  $\itemIndex ' \in \eta$ and positive ones $\itemIndex \in\ \Pi$ in the block and compute the ranking loss. This will result to the pairwise ranking loss, with respect to a block of item. We denote it $\ELoss (w) = \frac{1}{| \eta_\userIndex | | \Pi_\userIndex |}  \somme{\itemIndex ' \in\ \eta_\userIndex}{}{\somme{\itemIndex \in\ \Pi_\userIndex}{}{\loss_{\userIndex,\itemIndex, \itemIndex ',}(w)}}$. Applying the gradient descent algorithm on our values will result to : 
 $$ 
 \left[\begin{array}{c}
\UserRepr{\userIndex}{t + 1} \\ \\
  \ItemRepr{\itemIndex }{t + 1} \\ \\
  \ItemRepr{\itemIndex '}{t + 1}
\end{array}\right] = \left[\begin{array}{c}
\UserRepr{\userIndex}{t}\\ \\
  \ItemRepr{\itemIndex }{t}  \\ \\
  \ItemRepr{\itemIndex'}{t} 
\end{array}\right] - \begin{array}{c} 
\\ 
\alpha \times \nabla \ELoss (w)
\\ \\
\end{array}
$$

with $\nabla \ELoss (w) = \nabla  \frac{1}{| \eta_\userIndex | | \Pi_\userIndex |} \somme{\itemIndex ' \in\ \eta_\userIndex}{}{\somme{\itemIndex\in\ \Pi_\userIndex}{}{\loss_{\userIndex,\itemIndex, \itemIndex ',}(w)}} = \frac{1}{| \eta_\userIndex | | \Pi_\userIndex |} \somme{\itemIndex ' \in\ \eta_\userIndex}{}{\somme{\itemIndex\in\ \Pi_\userIndex}{}{ \nabla \loss_{\userIndex,\itemIndex,\itemIndex ',}(w)}} $ and\\ $\alpha \in \left[0, 1\right]$ the learning rate.

\chapter{Practical Implementation}
\begin{figure}
\begin{center}\includegraphics[width=.9\textwidth]{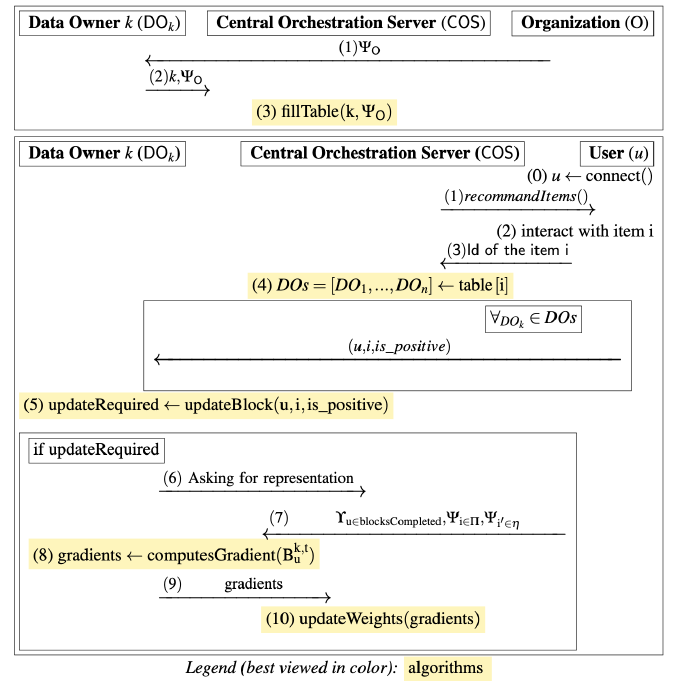}\end{center}
\caption{Workflow of DRIFT. The first diagram shows the preprocessing of the data, and
the second one shows the core of the algorithm The pseudo code of the algorithm fillTable,
updateBlock, computesGradient and updateWeights are available in Appendix 1, 2, 3 and 4
respectively.}\label{fig:41}
\end{figure}

Here we will focus on the practical implementation of DRIFT. We will first focus on the repartition
of the different values that it needs to compute in Section \ref{sec:41}, then we will see how the
gradients are computed in Section \ref{sec:42}. We will finish by seeing how the interactions are secure
in Section \ref{sec:42}. We provide a sequence diagram in Figure \ref{fig:41} which resume all the steps of
DRIFT.

\section{Repartition of the Task}
\label{sec:41}
Now that we know exactly how to compute our values, we need to disturb the computations among the participants. Recall that we have 2 major steps for updating the model, compute the loss, and update the global parameters.

Since these global parameters are contained in the $\COS$, our first idea was to send directly the completed block of interaction from each $\DO$ to the $\COS$. The principal problem with this method is that the $\COS$ can have access to all the interactions of the user, and this does not solve the problem of privacy. We need to hide these blocks from the $\COS$.

To solve that our second idea was to send directly the two vectors $\Items$ and $\Users$ to the $\DO$ when a block is finish. The principal problem with this method is that sending too much, and potentially useless, information could take a lot of time, and if multiples $\DO$ finish at the same it can create a bottleneck. Another problem with that method is that if our system is composed of only 2 $\DO$ each one is able guess which items are contained, and update in the other $\DO$, so this method does not solve the problem of privacy either. 

To solve that our third idea was to ask the $\COS$ to send only the representation of all items in blocks and the representation of the user. Each $\DO$ will compute directly the gradient following the gradient descent algorithm presented in Section \ref{sec:34}. Since the gradient value is lighter than the entire representation, the $\DO$ will send this result to the $\COS$, who will make the update on the global parameters. This method solves the problem of time complexity by sending only the information needed. Moreover it improves privacy by sending only needed information, and hides to the $\DO$ the information it doesn't need. But we still have a problem. The $\COS$ can save the old score vector for the user u, and compare it with the updated one. Thanks to these informations, after updating the model for a user, the $\COS$ can see if the items scores are increasing, decreasing or staying the same, and it can guess interactions present in the block of interaction. To solve this problem we add a threshold $\Theta$, such that the $\DO$ ask for an update if and only if the number of block completed is greater or equals than $\Theta$. We set $\Theta = 2$, thanks to that, the score of all the items with which users have completed blocks will be updated, and the $\COS$ can not conclude anything about the interaction with a specific user.

\section{Computation of the Gradients}
\label{sec:42}
Now that we know that the $\DO$ will compute and send the gradient of the loss function, we need to find a loss function and computes its gradients. In our experiments we will use the binary cross entropy loss = $- \frac{1}{N} \somme{i = 1}{N}{y_i  log(p(y_i)) + (1 - y_i)  log(1 - p(y_i))}$, with $y_i$ the label of the target, and $p(y_i)$ the probability that the target has the right label. In our case we will evaluate the loss between items in the completed block, so we already know which item is preferred, so we have a unique label equal to 1. We need to estimate the probability that the items are well classified, depending on w, as shown in Section \ref{sec:34}. We will compute this probability thanks to the sigmoid function on the score ($\sigma(w) = \frac{1}{1 + e^{-w}}$). We can conclude that in our case, we will have :
$$ 
\nabla \loss_{\userIndex,\itemIndex,\itemIndex'}(w) = \nabla -log(\sigma(w)) = - \frac{\nabla \sigma(w)}{\sigma(w)}
$$ \\
With : 
$$
\nabla \sigma(w) = \nabla \frac{1}{1 + e^{-w}} = \\
(\frac{1}{1 + e^{-w}}) ^2 \nabla (1 + e^{-w}) = 
-\frac{1}{1 + e^{-w}} \frac{e^{-w}}{1 + e^{-w}} (\nabla w) = -\sigma(w) (1-\sigma(w)) (\nabla w) 
$$
With 
$$
(\nabla w) = \left[\begin{array}{c}
\dfrac{\partial \User \times ( \Item -  \Item ' )}{\partial \User}\\
\dfrac{\partial \User \times ( \Item -  \Item ' )}{\partial \Item}\\
\dfrac{\partial \User \times ( \Item -  \Item ' )}{\partial \Item '}
\end{array}\right] = \left[\begin{array}{c}
( \Item -  \Item ' )\\ \\
\User \\ \\
-\User
\end{array}\right]
$$
This result to 
$$
\nabla \loss_{\userIndex,\itemIndex,\itemIndex'}(w)  = 
\left[\begin{array}{c}
(1-\sigma(w)) ( \Item-  \Item ' )\\ \\
(1-\sigma(w)) \User\\ \\
(1-\sigma(w)) (-\User)
\end{array}\right]  
$$

These 3 values will be computed by the $\DO$ and send to the $\COS$ who will update the global parameters, following the method described in Section \ref{sec:34}.

\section{Securing Interactions}
\label{sec:43}
Now that we know which data will be computed and sent, we need to know how to secure them. We will focus on the encryption of the triplets $\userInteract$. Recall that this one contains 3 informations, the user id, the item id, and a boolean saying if interaction is positive or not. This triplet will be shared between the user and several $\DO$ that have the item. Since this interaction can share sensitive information, and since another participant can intercept this message, we need to secure this exchange to avoid harming the user's privacy. To do that we will use AES-GCM, presented in the state of the art (Section \ref{sec:24}). At the reception of an interaction the $\DO$ will decrypt it, thanks to the key share when it was created. To optimize our architecture we will not encrypt all the triplets, but only the user id and the item id, as a unique tuple. This grants about 50\% less time and since a lot of interaction will be made, this gain is significant.

\chapter{Theoretical Analysis}
\label{chap:5}
We analyse the convergence of DRIFT in Section \ref{sec:51}, the security of DRIFT in Section \ref{sec:52}, and the complexity of DRIFT in Section \ref{sec:53}.
\section{Convergence}
\label{sec:51}
We recall that the preprocessing of the data consists of splitting the item into K $\DO$, then each item of a dataset S is present in at least one $\DO$, so at each step, at least one block will be updated. All the interaction between a user and an item will be done directly in the corresponding $\DO$. We can conclude that all the interactions of a dataset S, will be split into smaller datasets ($\mathrm{S_1, S_2, ... , S_K}$). Each of them will participate in the update of the global parameters, by adding his computed loss, we can conclude that the final loss of our algorithm will be : $\mathcal{L}(w)$ = $\sum_{i=1}^{M}{\mathcal{L}_i(w)}$ , with $\mathcal{L}_i(w)$ the total loss created by the $\DO_i$. We know that the time taken by the $\DO$ to compute the gradient, depend on a finite block of interaction, so we know that there exist D < $\infty$, such that for any $\DO_i$, the delay $d^i_k$ < D. Assume that each $\mathcal{L}_i$ is differentiable and that  $\nabla \mathcal{L}_i$ is $\frac{1}{L}-cococercive$. But since the delay is bounded and assumptions on the loss function hold, the sequence produced by our Distributed Gradient update rule converges to the unique minimizer. (Theorem 1 \cite{JIA2018}). Moreover, in our experiments, we will use the same loss function as SAROS, we can conclude that both algorithms should converge to the same minimizer.

\section{Security}
\label{sec:52}
\begin{figure}
\begin{center}
\includegraphics[width=.8\textwidth]{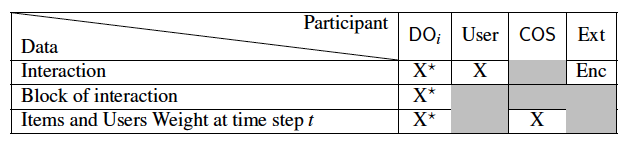}
\caption{Security properties of DRIFT. The X means that the participant can see in clear the
informations. X$^\star$ means that the DOi can see in clear the informations only if it have the items
involved. Ext means an external observer having access to all messages exchanged between
participants. Enc means that the data are visible but encrypted with AES-GCM. A grayed cell
means that the participant cannot see the information.}\label{fig:51}
\end{center}
\end{figure}

The safety characteristics of $\proto$ are summarized in Figure \ref{fig:51}. We will show that all participants cannot guess the user interacts with a probability better than random. 

\paragraph{Security of the user with respect to $\DO$.}
Thanks to the architecture of $\proto$, each $\DO$ knows data about its items, both items and blocks are private, so they cannot have any information about another $\DO$. 
This means that at a timestamp each $\DO$ cannot guess the interactions a user has made with another $\DO$. Moreover, since each $\DO$ has a different key for decrypting the interactions, they cannot guess interactions that do not concern it with a probability better than random.

\paragraph{Security of the user with respect to the $\COS$.} \label{sec:COS}
Thanks to the architecture of $\proto$, the $\COS$ only knows about the global parameters. It will receive the gradient to update the representation of several items and users, depending on the hyper-parameter $\Theta$. Moreover, unlike most recent works where the users send their gradients to the server, our algorithm shares the gradients from the $\DO$ to the $\COS$, so even if the $\COS$ has access to the scores of items for users, the $\COS$ in unable to conclude which interaction was made by which user. We can conclude that the $\COS$ cannot guess any user interactions, with a probability better than random. 

\paragraph{Security of the user with respect to an $\mathsf{External\ observer.}$}

Thanks to the architecture of $\proto$, there is only 4 information that can be sent on the network, interactions, item representations, user's representations and the gradients computed by $\DO$s. \\
Interactions are encrypted thanks to AES-GCM, which is IND-CPA secure, so an external observer cannot guess the interaction with a probability better than random.\\
But other values are not encrypted, so the external observer will have access to items representations, users representations and the gradients computed by $\DO$s. We can conclude that it does not have more information than the $\COS$. Following the same principles as \ref{sec:COS}, we can conclude that an external observer cannot guess the interactions made by users, with a probability better than random.

\section{Complexity}
\label{sec:53}
We will study here the complexity of $\proto$.  
Since the preprocessing is done once at the beginning, the time spent in this part is constant, we will not take it into account in our results. \\
At each timestamp 3 operations can be done. We need first to secure and send the interaction, then to share the parameters, and update the model. We will first assume that at each timestamp, only one $\DO$ participate. \\
First, we will focus on the sending of the user interaction, this is our only cryptographic interaction. Denote A, the time complexity to encrypt and decrypt a tuple composed of the user id and the item id, with AES-GCM. $\proto$ requires $O(N)$ encryption, which means that the time to share the information between the user and the $\DO$ is on O(NA). 

Sharing the parameter consists of sharing raw information between the $\DO$ and the $\COS$. Since we share only information for a block of interaction, the size of these two messages is bounded by this block's size. We assume that there is no perturbation in the network, so we assume that the time spent for these communications is negligible. \\
The last operation made by our algorithm is the update of the global parameters. Each value will be updated after receiving the gradients, the total complexity of this operation will be on $O(N)$.\\
The total complexity of our algorithm will then be on $O(NA + N)$. Suppose now that our system is composed of K $\DO$, and that at each timestamp several $\DO$ participate. Since they make the same operation, we can conclude that the total complexity of $\proto$ will be on $O(K(NA + N)) = O(NA)$, which is linear in $N$.

\chapter{Experimental Performance of the Solution}
In this chapter we will present the results of our experiments. We present first the experimental
evaluation (Section \ref{sec:61}), where we present the metrics used, and the experimental setting. Then
we analyze the quality of the recommendations (Section \ref{sec:62}). We will finish with an analysis
of the repartition of the time during the training of DRIFT. (Section \ref{sec:63}).
\section{Experimental Evaluation}
\label{sec:61}

\subsection{Metrics}
\label{sec:611}
We will focus on 2 metrics based on the precision of our model, in other words, based on the number of relevant items recommended. First the MAP as $\frac{1}{N} \sum_{i=1}^{K}{AP_k(u)}$ with $AP_k(u)$ The average precision for the user u. \\
Then the NDCG = $\frac{DCG}{IDCG}$. Denote the DCG the Discounted Cumulative Gain, as $\sum_{i=1}^{K}{\frac{2^{rel_i}-1}{log_2(i + 1)}}$, with $rel_i$ equals to 1 if the item is relevant, 0 otherwise. Denote the IDCG the Ideal Discounted Cumulative Gain, which correspond to the maximum possible value for the DCG, when all the items are pertinent, $\sum_{i=1}^{K}{\frac{1}{log_2(i + 1)}}$. 

\subsection{Experimental Setting}
\label{sec:612}
We aim to compare our algorithm with SAROS, to see how both algorithms evolve. As we said in Section \ref{sec:51}, we will use the same loss function for both algorithms. \\
This comparaison will be made using a dataset that contains user rating for movies i.e. using MovieLens \cite{HarperK16}. The preprocessing part is made by splitting movies per genre, and creating one $\DO$ for each. If a movie has multiple genres, it will be contained in multiple $\DO$. \\
We will next focus on the interaction of users. As we need to have implicit interaction, and since the dataset gives us grades on the movie (from 1 to 5), we have considered that the interaction is negative if the rate is lower than 3, otherwise we have considered it as positive. We also need to keep the context to build the blocks of interaction, to do that we sort the dataset by timestamp. Then we keep 80\% first user interactions for the training part, and we will try to guess his 20\% remaining interactions, which represent our testing part. \\
We did our experiments on a virtual machine running Ubuntu, located in a server with a 32 cores Intel(R) Xeon(R) CPU E5-2640 v3 @ 2.60GHz and a nvidia RTX A6000. Since we also need to focus on the efficiency of our algorithm, we launched SAROS and $\proto$ at the same time to avoid any perturbation of the server. Moreover to have a more efficient algorithm we will also use tensorflow for the update and for the recommendation. Each $\DO$ will compute the loss and the gradient of this one, thanks to the gradients computed in Section \ref{sec:42}. We also add a ridge regression to our loss function, which makes it strongly-convex, and respect all the assumption presented for the correctness, shown in Section \ref{sec:51} \\ 

\section{Analysis of the Recommendation}
\label{sec:62}
We will now focus on the recommendation of our system, compared to the SAROS ones. First, we will focus on the evolution of the loss during the training (Section \ref{sec:621}), then we will focus on the result of our metrics, presented in the Section \ref{sec:611} (Section \ref{sec:622}).

\begin{figure}
\begin{center}
\includegraphics[width=.8\textwidth]{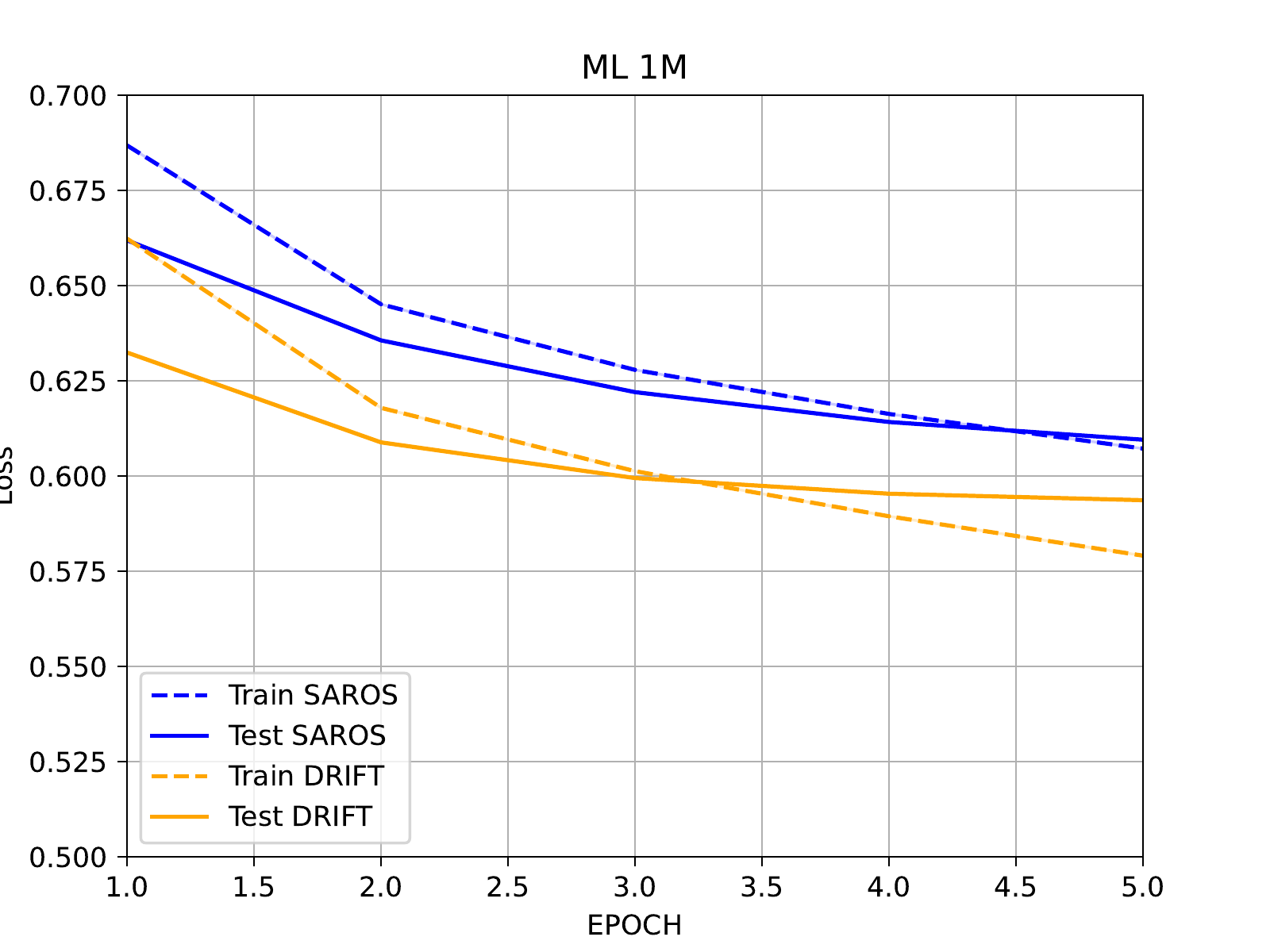}
\caption{Evolution of the loss during the training and testing of SAROS and DRIFT}\label{fig:61}
\end{center}
\end{figure}

\subsection{Evolution of the Loss}
\label{sec:621}
We will now focus on the evolution of the loss during the training part and the polyvalence of the algorithm, showing as the testing part. 
The result of the experiment is available on the figure \ref{fig:61}. As we can see the evolution of the algorithm is very similar. But we can see that at the beginning $\proto$ has a lower loss than SAROS, but converges after few epochs.  

\subsection{Evolution of the Metrics}
\label{sec:622}
\begin{figure}[t!]
\begin{center}
\begin{tabular}{cc}
\includegraphics[width=.45\textwidth]{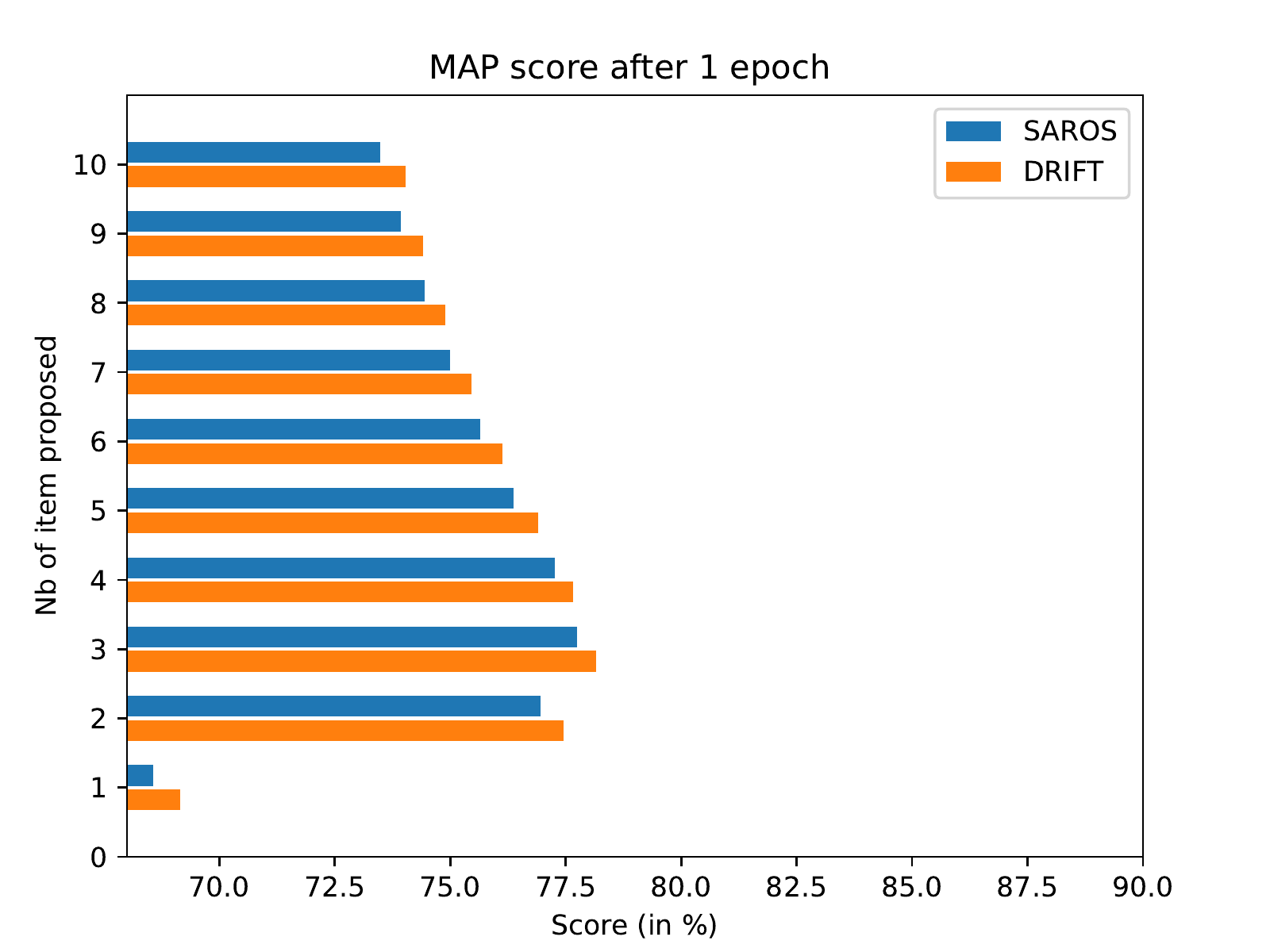} & \includegraphics[width=.45\textwidth]{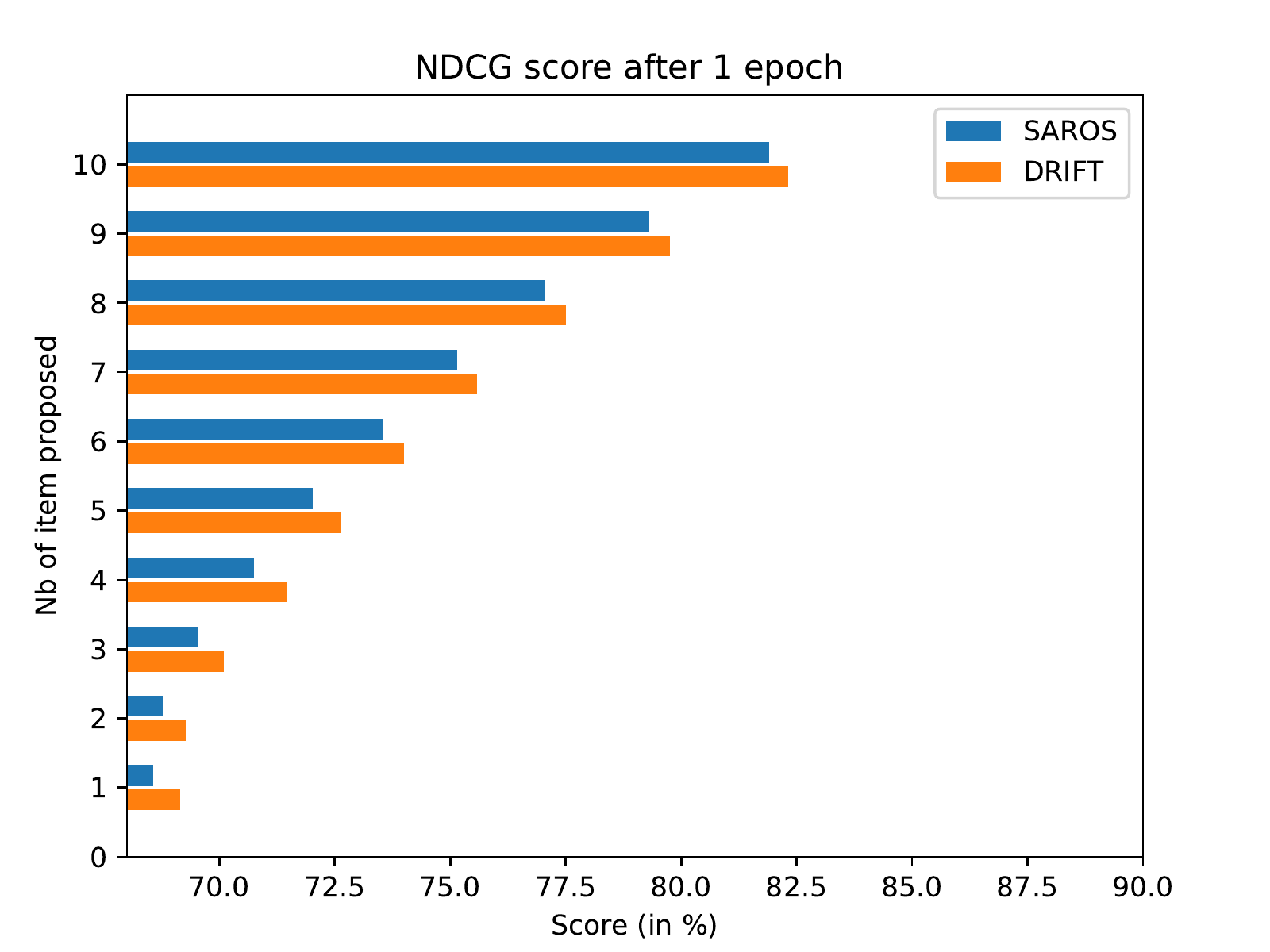} \\
\includegraphics[width=.45\textwidth]{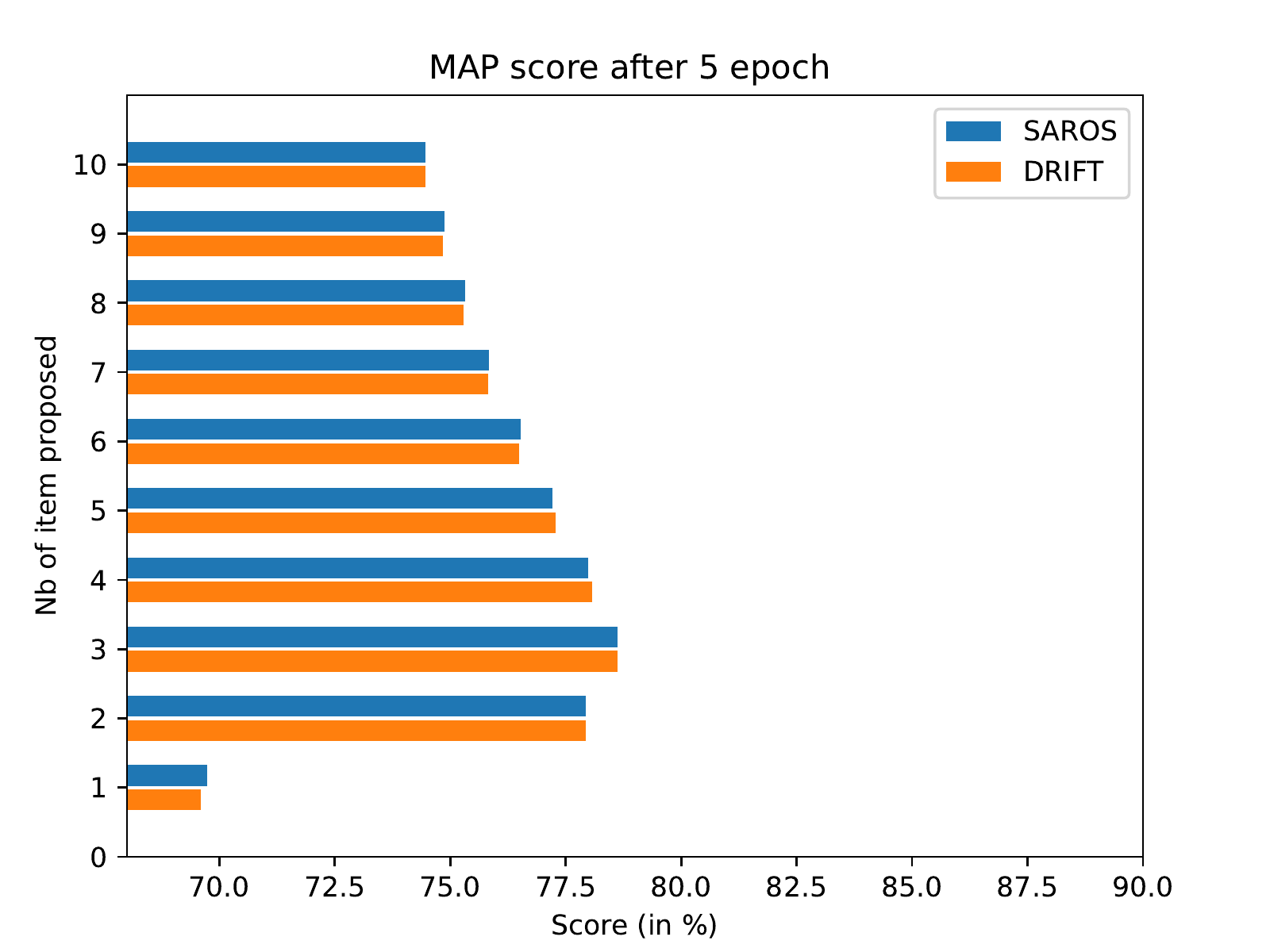} & \includegraphics[width=.45\textwidth]{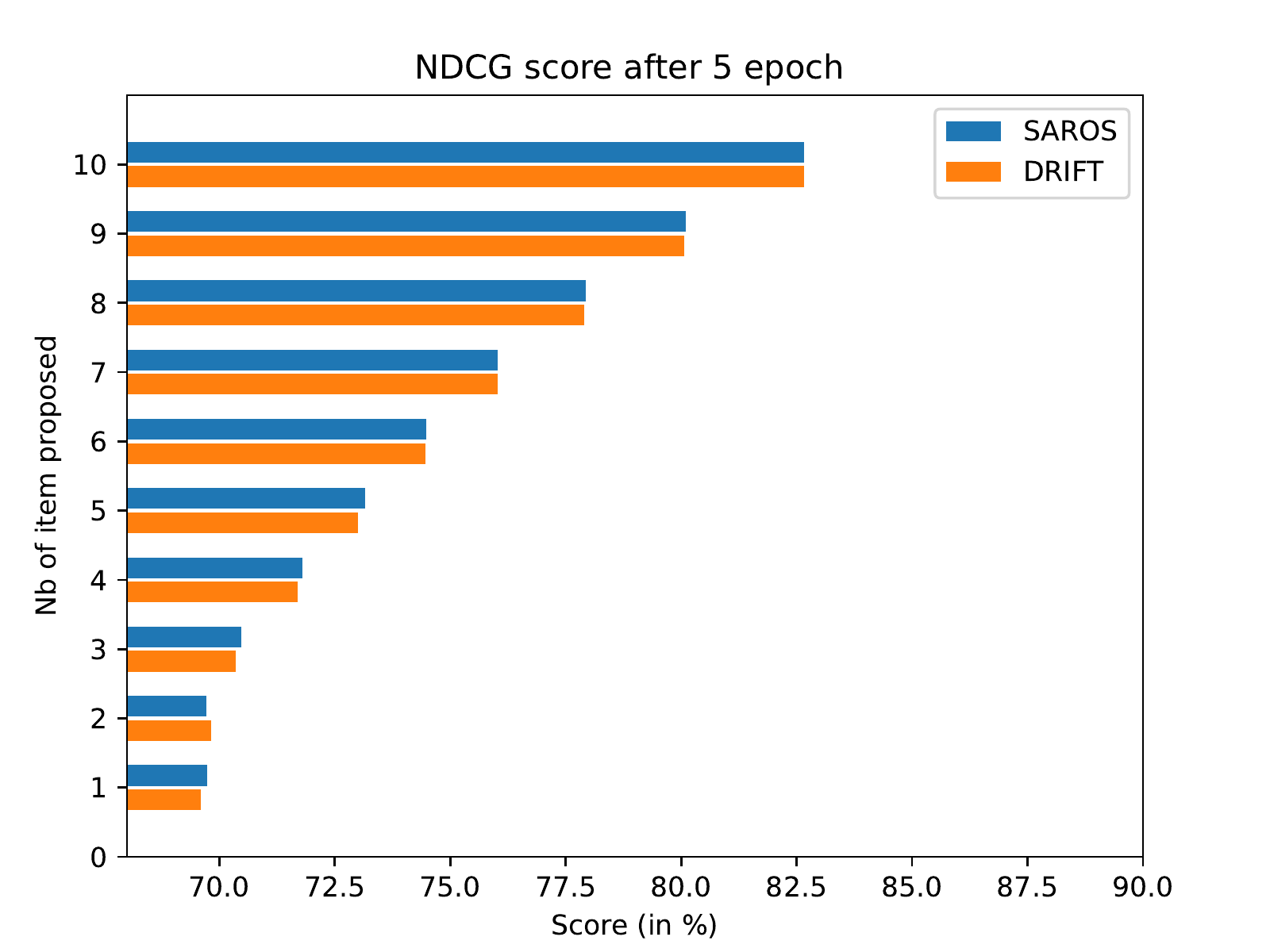}
\end{tabular}
\caption{Evolution of the Recommendation after a single Epoch.}\label{fig:62}
\end{center}
\end{figure}

Thanks to the observation of the loss made in Section \ref{sec:621}, we will focus on the recommendation at the very beginning, and at the end of the training. The results of our metrics are available in Figure \ref{fig:62}. We observe that after one epoch, our architecture outperforms SAROS. We thought that this is due to the preprocessing. In our dataset there is an average of 3 genres per movie, so we can conclude that each block created in SAROS will result in an average of 3 blocks in $\proto$. Theses blocks will contains less negative items, and will update multiples time the score of the items, this can explain why our recommendation are more precise. To continue to observe this phenomenon, we also compare the results after training. We observe that after the training our architecture does not evolve a lot, unlike SAROS who evolve a lot. This is due to the high number of blocks in $\proto$ that made it converge earlier than SAROS. However we observe that at the end, our metrics are very close, which confirm the correctness of the algorithm, shown in Section \ref{sec:51}. 

\section{Analysis of the Time}
\label{sec:63}
\begin{figure}[!]
\begin{center}
\includegraphics[width=.45\textwidth]{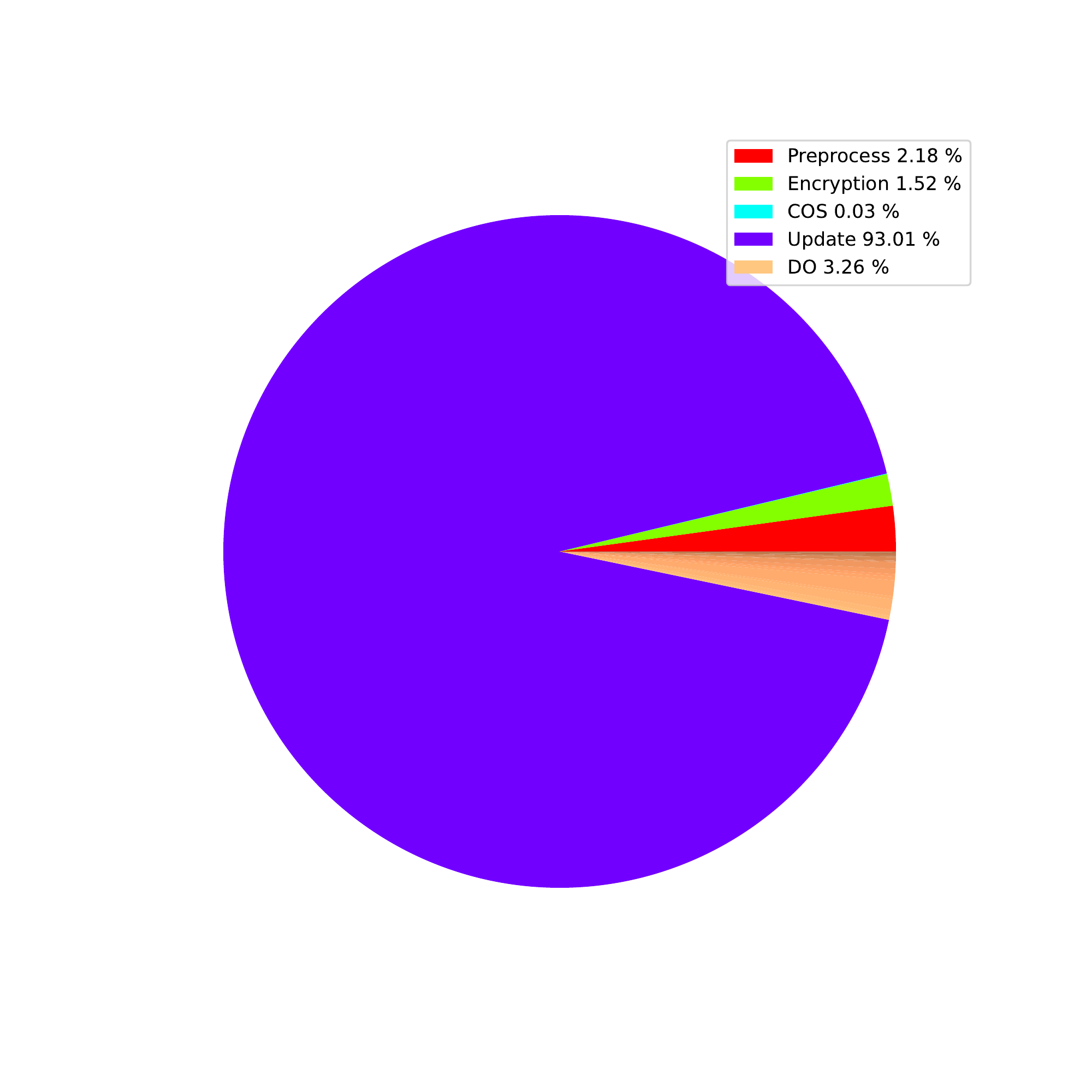} 
\caption{Evolution of the time during the running of one epoch. The non label part representing
one Data Owner. The legend gives the sum of time spend into the different Data
Owner.}\label{fig:63}
\end{center}
\end{figure}

We provide the details of the total time of computation at figure \ref{fig:63}. As we can see, we spend the most of the time in the update part. Less than 10\% of the time is used for the security layer of the algorithm, which is reasonable, especially when some part, such as the preprocessing, are constant. Thanks to the fact that we optimize the number of values encrypted, and thanks to the efficiency of AES-GCM, the encryption takes only 1.52\% of the total time. 

\chapter{Summary of results, Conclusions, Expected Impact}
In this report we aim to solve the problem of security in recommender systems, with a federated recommender system algorithm using implicit feedback. The main piece of our architecture are the $\DO$s which manage the interaction of the user, and participate in the update of the model by communicating with the $\COS$. We proposed $\proto$, a federated recommender system with implicit feedback on the items. This algorithm is shown to be very efficient, more than 90\% of the time is dedicated to the update of the global parameters. 

Thanks to a theoretical analysis of the security, where we demonstrated that each participant cannot guess the interactions of the users with a high probability, we showed that the privacy is preserved. We also provide a proof of convergence that shows that our algorithm has the potential to learn as well as a non secure version. Moreover, thanks to our experiments, we can see that our algorithm is shown to be as precise as a non secure version, on MAP and NDCG measure. 
\paragraph{Future work} 
As future work we aim to test DRIFT in real federated settings, with multiple
machines, each representing one DO and the COS. We also aim to dynamically manage the
number of DOs and users in the system, to see how the recommendation will evolve over time
on larger datasets \cite{KASSANDR,PANDOR}. It will be also needed to focus on the minimal number of completed
blocks before update, focusing on our hyper-parameter $Q$, if this one is too small, there will
be a lot of communication, who can bring some bottleneck, If this one is too large the global
parameters will be updated only slightly, which may cause bad recommendations for users in a
semi-supervised scenarios \cite{JAIR,AAAI} where all user feedback are not available.
We also aim to test DRIFT on different datasets to complete our results. The main problem
with the training is the repartition of data into different datasets.

\appendix \chapter{Appendix}


\begin{algorithm}[t]
\caption{Used by the COS during the preprocessing at step 3}\label{alg:fillTable}
\begin{algorithmic}
	\Require \textbf{int} k, \textbf{ List[int]}\ items
	\For{item $in$ items}
		\If {$not\ self.table.contains(item)$}
			\State $self.table[item] \gets [\ ]$
				\EndIf
			\State $self.table[item].append(k)$
	\EndFor
\end{algorithmic}
\end{algorithm}


\begin{algorithm}[t]
\caption{Used by the $\DO$ to update his blocks at step 6}\label{alg:updateBlock}
\begin{algorithmic}
\Require $\textbf{ int}\ u, \textbf{ int}\ i, \textbf{ bool}\ is\_positive,$ 
\State $isFull \gets False$ 
\State $\Block{k}{t}{u} \gets self.getBlockForUser(u) $ 
\State $\Negative{u}{t} \gets  \Block{k}{t}{u}.negativeInteractions$ 
 \State $\Positive{u}{t} \gets \Block{k}{t}{u}.positiveInteractions$ 
\If {is\_positive}
    \State $\Positive{u}{t}.add(i)$ 
\ElsIf{$\Positive{u}{t}.isEmpty()$}  
    \State $\Negative{u}{t}.add(i)$ 
\Else 
		\State self.save[u].append($\Block{k}{t}{u}) $ 
		\State $\Block{k}{t + 1}{u} \gets newBlock$ 
		\State $\Block{k}{t + 1}{u}.negativeInteractions.add(i)$
		\State $updateRequires \gets self.save.size \ge \Theta $ 
\EndIf \\
\Return $updateRequires$
\end{algorithmic}
\end{algorithm}


\begin{algorithm}[t]
\caption{Used by the $\DO$ to computes the gradient thanks to the representation of the items at step 8}\label{alg:grad}
\begin{algorithmic}
	\Require \textbf{List[List[double]]} $\Items^+$, \textbf{List[List[double]]} $\Items^-$, \textbf{List[List[double]]} $\Users$
	\For {u $\textbf{in}$ self.save } 
	\State $b \gets self.save[u]$
	\For{n $\textbf{in}$ $b.negativeInteractions$ } 
		\For {p $\textbf{in}$ $b.postiveInteractions$}
			\State $d\_u, d\_ipos, d\_ineg \gets \nabla LossFunction(\UserRepr{u}{t}, \ItemRepr{n}{t}, \ItemRepr{p}{t})$
			\State $gradients\_items.append((p, d\_ipos))$
			 \State $gradients\_items.append((n, d\_ineg))$ 
			\State $gradients\_user.append((u, d\_u))$ 
			
		\EndFor
	\EndFor
	\EndFor
	\State $self.save.reset()$\\
	\Return $gradients$
\end{algorithmic}
\end{algorithm}


\begin{algorithm}[t]
\caption{Used by the $\DO$ to update the weight matrix thanks to the blocks at step 8}\label{alg:updateWeights}
\begin{algorithmic}
	\Require $\textbf{List}\ gradients\_user \textbf{List}\ gradients\_item$
	\For {$u, gradient$ $\textbf{in}$ $gradients\_user$}
		\State $\UserRepr{u}{t+1} \gets \UserRepr{u}{t} - \alpha \times gradient$
	\EndFor
	\For {$i, gradient$ $\textbf{in}$ $gradients\_item$}
		\State $\ItemRepr{i}{t+1} \gets \ItemRepr{i}{t} - \alpha \times gradient$
	\EndFor
	
\end{algorithmic}
\end{algorithm}

\newpage
\backmatter

\bibliographystyle{plain} 
\bibliography{references.bib}
\end{document}